\documentclass{article}
\usepackage[utf8]{inputenc}
\usepackage{graphicx}
\usepackage{blindtext}
\usepackage{color}
\usepackage{hyperref}

\title{Ground state dynamically stable phases for fluorine in the TPa pressure regime by evolutionary algorithms}
\author{B. H. Cogollo-Olivo$^{a,b,*}$ and J. A. Montoya$^{a}$}
\date{$^{a}$Universidad de Cartagena, Instituto de Matem\'aticas Aplicadas, 130001, Cartagena de Indias, Colombia; $^{b}$Universidad de Sucre, Doctorado en Ciencias F\'isicas, 700001, Sincelejo, Colombia}

\begin{document}

\maketitle

\begin{abstract}
In this work, we employed \textit{ab initio} methods combined with evolutionary algorithms for searching stable structures for fluorine in the terapascal (TPa) regime. We performed several structural searches using the USPEX code, at pressures that spanned from 1 to 5 TPa and considered up to 16 atoms per cell for selected pressures. Our findings partially support recent studies by validating the transformation of fluorine from a molecular form, $Cmca$, into an intermediate polymeric form before its eventual dissociation. In fact, the enthalpy comparisons between candidate structures of fluorine at high pressure show a direct transition from the molecular phase $Cmca$ into a $Pm\bar{3}n$ extended structure at 2.7 TPa, the later consisting of linear chains and independent atoms, which disagrees with previous conflicting reports that proposed other intermediate phases to also exist as stable crystalline forms close to 3 TPa. 
\end{abstract}

\section{Introduction}
Under extreme conditions of pressure and temperature, the chemistry of materials, as we know it, changes drastically, making it possible to access new states of matter. Thus, the lab and theoretical research under extreme conditions have been fundamental in their contribution to geophysics and mineralogy, as well as to the characterization of new materials of technological interest and their respective phase diagrams, leading to a better understanding of their physical-chemical properties. In this sense, the study of diatomic molecules of elements with low atomic numbers subjected to high-pressure conditions, has been the subject of great interest by the scientific community due to the theoretical and experimental detection of phenomena, such as the eventual metallization induced by pressure and phase transformations in analogous mono-atomic systems such as nitrogen \cite{Nellis-1984,Reichlin-1985} and the halogen elements \cite{Fujii-1989,Takemura-1980,vanBolhuis-1967,Duan-2007}. Given its similarities with hydrogen and the high energy densities of its electronic bonds, fluorine was chosen as the study object of this work. This element is the first of the halogens and the last of the diatomic elements of the first row; likewise, it is an abundant element in the universe and is particularly interesting due to its high oxidizing capacity and the highest electronegativity among all elements, followed by oxygen's.

However, fluorine has not been studied extensively compared to other single-element diatomic molecules (H$_{2}$, N$_{2}$, O$_{2}$, Cl$_{2}$, Br$_{2}$, and I$_{2}$). For instance, the space group of solid fluorine at high pressure was controversial for many years because two different structures were proposed through spectroscopy studies and theoretical calculations, respectively \cite{Gamba-1987,Kobashi-1980,Kirin-1986,Schiferl-1987}. The experiments were carried out using diamond anvil cells (DAC), but it was challenging to obtain reliable results \cite{Schiferl-1987,Pravica-2014}. This technical difficulty is because fluorine and other halogen elements are characterized by being volatile, corrosive, and highly reactive, increasing the complexity level when studied experimentally. For this reason, and using computational techniques, it was not until a few years ago that a study established that the $C2/c$ structure is more stable than $C2/m$ and then transforms into $Cmca$ at 8 gigapascals (GPa) \cite{Lv-2017} without reporting any other structural transformation up to the pressure-limit of that work, which was 100 GPa. In their study, Olson \textit{et al.} \cite{Olson-2020} spanned a higher pressure range covering up to a few terapascals (TPa); not only they placed the transition from $C2/c$ to $Cmca$ at 70 GPa, but they also reported phase transitions to $P4_{2}/mmc$ and then to $Pm\bar{3}n$, at 2.5 and 3.0 TPa, respectively. Additionally, the study by Olson \textit{et al.} \cite{Olson-2020} includes a structural search component: their work makes use of a recently refined methodology called Symmetry Driven Structure Search (SYDSS) \cite{Domingos-2018}; the method is based on sampling space groups and Wyckoff positions to propose new crystal structures. In Ref. \cite{Olson-2020}, the structural search for high-pressure phases received as an starting point a set of structures previously reported for low pressures, while some restrictions were imposed to generate new systems, \textit{i.e.}, the bond lengths and the number of atoms in the molecules. More recently, Duan \textit{et al.} \cite{Duan-2021} presented enthalpy relations for predicted phases for fluorine up to 30 TPa based on the \textit{ab initio} random structure searching (AIRSS) method \cite{Pickard-2011}. The structural sequence reported in that work undergoes two intermediate transitions, namely $P6/mmc$ and $Pm\bar{3}n$, at 2.7 and 4.0 TPa, respectively, before transforming into the $Fddd$ atomic phase at 30 TPa. 

Given the substantial differences observed in recent theoretical reports on high-pressure sold fluorine, validating the structures found in Refs. \cite{Olson-2020} and \cite{Duan-2021} is necessary. For this study, we chose the computational tool known as USPEX \cite{Oganov-2006,Glass-2006,Oganov-2011,Lyakhov-2013}, which is currently one of the most reliable codes used within the materials science community for crystal structure determination, that combines \textit{ab-initio} methods with structure prediction techniques based on evolutionary algorithms. Among its many applications, it has been able to predict new phases of materials at high pressure, including those relevant to the Earth's interior, allowing new ways of understanding matter at extreme conditions.

In this work, we performed a structural search for fluorine varying the number of atoms per unit cell at different volumes with the USPEX code interfaced with Quantum ESPRESSO \cite{Giannozzi-2009,Giannozzi-2017} as the selected tool for accurate \textit{ab-initio} structural optimization, as detailed in the Methods section. The Results and Discussion section provides this investigation's findings and analysis for several structure-candidates. Finally, the article concludes with a summing-up in the Conclusions section, that provides answers about the type and range of stability for fluorine's stable phases, in the pressure range from 1 to 5 TPa.

\section{Methods}

The main objective of this work is to provide independent insight using a different approach to performing a structural search for novel phases of fluorine in the terapascal regime. Over the past years, the likelihood of the reliability results obtained with computing methods for crystal structure prediction has been improved substantially. In this work, we have selected the USPEX package for its proven robustness in the pressure regime of this study. The initial stage used USPEX to successfully generate good candidate structures for solid fluorine, followed by density functional theory (DFT) calculations for structural optimization of selected phases at different pressures. In addition, the density functional perturbation theory (DFPT) scheme allowed us to determine the dynamical stability of the structures found, from vibrational data.

\subsection*{USPEX Calculations}
USPEX employs an evolutionary algorithm to generate different structures based only on the number and type of atoms within the unit cell \cite{Oganov-2006,Glass-2006,Oganov-2011,Lyakhov-2013}. At first, the initial set of structures is generated by randomly choosing any crystal space group. Thus, this first generation is mainly composed of far-of-equilibrium structures. It is then necessary to perform a structure relaxation to determine the total internal energy of each candidate system and use that as a fitness criterion. After a purely random start, new generations are created via genetic operations which can apply criteria such as heredity, mutation, and permutation; nevertheless, a portion of the new structures is still made randomly to ensure a level of diversity at each generation. The structural search ends when one of these criteria is achieved: No new \textit{best structure} is generated for a certain number of generations, or, The set value for the maximum number of generations is reached.

Here, fixed composition searches with different atoms per cell were conducted to determine suitable high-pressure phases for solid fluorine. USPEX-based structure predictions were performed independently on different cell sizes, containing 2, 4, 6, 8, 12, and 16 atoms, in a pressure range spanning from 1.0 to 5.0 TPa in steps of 0.5 TPa, with the exception of 1.5 TPa that was not considered explicitly in this work. Also, the two biggest cell sizes were studied at four selected pressures instead of eight in that same range. The calculations were performed with an initial population of 30 structures generated randomly, with no restrictions on the space groups that can be assigned to the structures. In addition, no seeds or anti-seeds were considered. Subsequent generations were produced by applying the following variation operations (in parenthesis are the values used in this work): heredity (50\%), soft-mutation (15\%), lattice mutation (10\%), and permutation (5\%), while the rest of the newly generated structures (20\%) were still randomly produced with the random symmetric structure generator. We set the maximum number of generations to 50, and the convergence criterion was set to 20. All USPEX calculations ended by reaching the \textit{best structure} convergence criterion.

\subsection{DFT Calculations}
The trial structures were optimized in a low-to-high precision process using first-principles DFT calculations using the Quantum Espresso suite \cite{Giannozzi-2009,Giannozzi-2017} for variable-cell relaxation. We increased the convergence criteria and precision settings at each step, with a reciprocal space resolution for k-point generation of 0.20, 0.16, and 0.12 2$\pi$/\AA, respectively. The k-point grids for the Brillouin zone integration were generated with the Monkhorst-Pack method \cite{Pack-1977}. Finally, all DFT calculations were performed using the plane-wave pseudopotential method with projector-augmented wave (PAW) pseudopotentials \cite{Blochl-1994}. For the calculation of the exchange and correlation functionals, generalized gradient approximation (GGA) combined with Perdew–Burke–Ernzerhof (PBE) \cite{Perdew-1996} parametrization was used. With the calculated lattice enthalpy after total relaxation, the structures were sorted from the lowest to the highest to be later used as input for the next generation.

\section{Results and Discussion}
\label{sec:examples}

We began our exploration keeping first in mind the structures that were reported in previous studies up to 5.0 TPa, for future reference and comparisons \cite{Lv-2017,Olson-2020,Duan-2021}]: $Cmca$, $P4_{2}/mmc$, $P6/mcc$, $Pm\bar{3}n$. Their structural parameters are detailed in table \ref{table1}.

\begin{table}
 \centering
 \caption{Structural information of previously reported structures for high-pressure phases of fluorine.}
 \begin{tabular}{c c c}
  Space group & Lattice parameters & Atomic coordinates\\
  Pressure & (\AA, °) & (fractional) \\
  \hline
  (64) $Cmca$        & $a = 4.31$                     & $8f$ \\
  0.02 TPa           & $b = 2.88$                     & $y = 0.3799$ \\
  Ref. \cite{Lv-2017}     & $c = 5.84$                     & $z = 1.3933$ \\
                     & $\alpha = \beta = \gamma = 90$ &  \\
  (131) $P4_{2}/mmc$ & $a = 2.69$                     & $2a$ \\
  2.75 TPa           & $b = c = 2.60$                 & $2d$ \\
  Ref. \cite{Olson-2020}  & $\alpha = \beta = \gamma = 90$ &  \\
  (192) $P_{6}/mcc$  & $a = 3.68$                     & $2a$ \\
  3.00 TPa           & $b = c = 2.65$                 & $12l$ \\
  Ref. \cite{Duan-2021}   & $\alpha = \beta = 90$          & $x = 0.1097$ \\
                     & $\gamma = 120$                 & $y = 0.6831$ \\
  (223) $Pm\bar{3}n$ & $a = 2.59$                     & $2a$ \\
  3.15 TPa           & $\alpha = \beta = \gamma = 90$ & $6d$ \\
  Ref. \cite{Olson-2020}  &                                &  \\
  5.00 TPa           & $a = 2.44$                     & $2a$ \\
  Ref. \cite{Duan-2021}   & $\alpha = \beta = \gamma = 90$ & $6c$ \\
 \end{tabular}\label{table1}
\end{table}

\subsection{Structural search process}
In total, nearly 30000 fluorine structures with 2-16 atoms per unit cell spanning the pressure range of 1 to 5 TPa were generated in 40 different runs, via USPEX combined with DFT, using the Quantum ESPRESSO suite. The first generation of each run consists of randomly generated structures. After that, the act of creating a new candidate structure usually follows a two-step procedure. First, the evolutionary algorithm enables the generation of a new system using one of three operations \cite{Oganov-2006,Glass-2006,Oganov-2011,Lyakhov-2013}:
\begin{itemize}
  \item \textit{Heredity} combines two or more parent structures to create a new one, allowing a broad search of the energy landscape while preserving fragments of good structures.
  \item \textit{Permutation} randomly swaps chosen pairs of atoms within the unit cell, which facilitates finding the optimal ordering of the atoms.
  \item \textit{Mutation} creates a child structure from a single parent, retaining some characteristics from it while introducing new structural features. There are two main types of mutations: \textit{lattice mutation}, that randomly distorts the cell shape, allowing a better exploration of the neighborhood of parent structures and preventing premature convergence of the lattice; and \textit{softmutation}, which moves the atoms along the eigenvectors of the softest modes in order to create a new structure across the lowest energy barrier, allowing a local and semi-local exploration of the energy landscape.
\end{itemize}

\begin{figure}[ht]
\centering
\includegraphics[width=\linewidth]{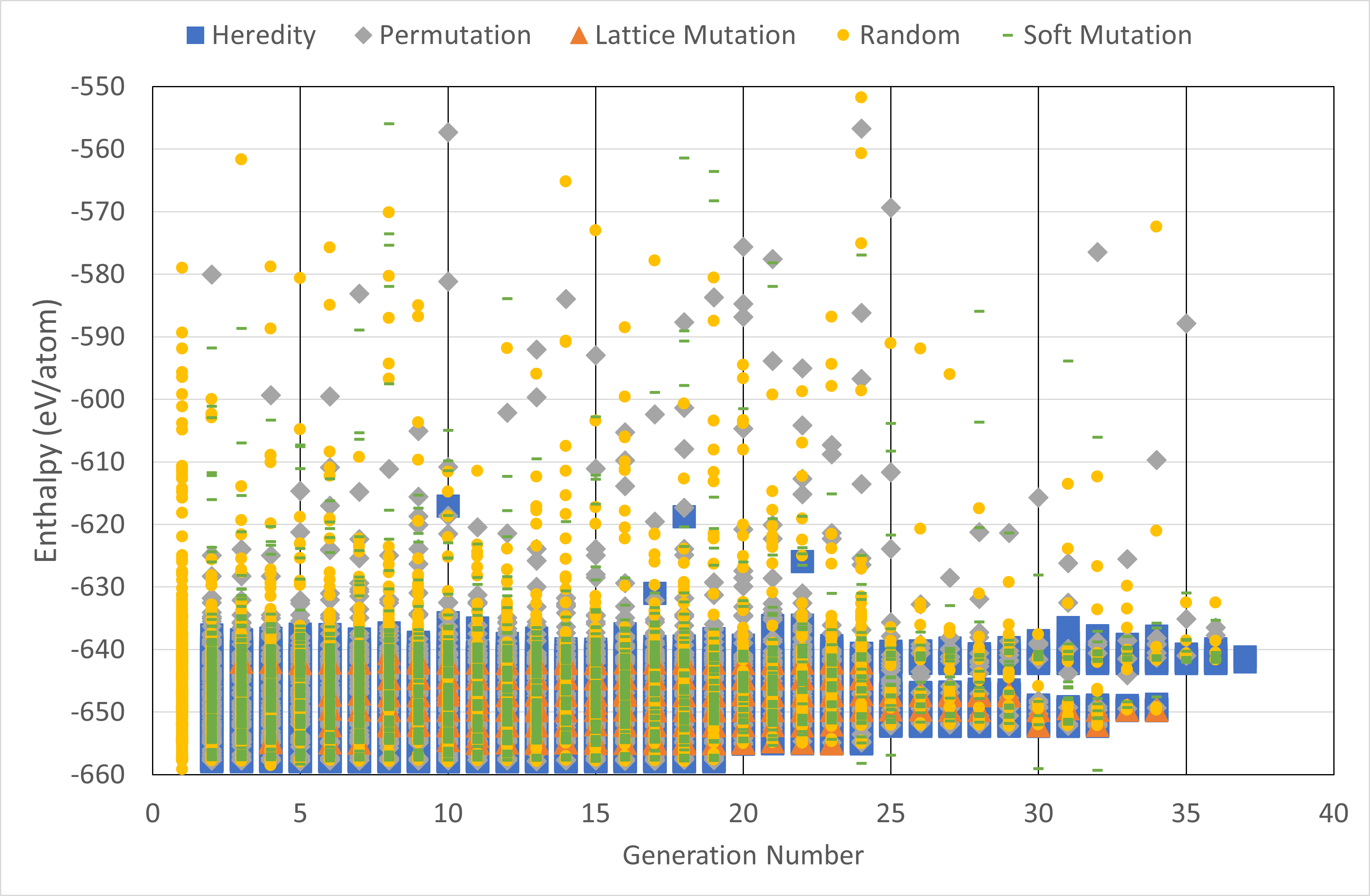}
\caption{Enthalpy distribution, in eV per atom, of structures generated during 40 different runs for fluorine with 2-16 atoms per unit cell and at 1-5 TPa. The symbols represent the origin of generated structures: random generation (yellow circles) or as a result of certain genetic operations: heredity (blue squares), permutation (grey rhombus), lattice mutation (orange triangles), and soft mutation (green dash).}
\label{figure1}
\end{figure}

Each run contemplates the usage of the genetic operations described above, together with a percentage of randomly generated structures. Figure \ref{figure1} shows the enthalpy distribution of the generated systems within all 40 runs performed. We can see that randomly generated structures (yellow circles) with those from permutation operations (grey rhombus) and up to some extent also the ones from soft mutation (green dashes) span a large energy interval. In contrast, most relevant structures are created due to heredity (blue squares) and lattice mutation (orange triangles).

\subsection*{Crystal structure determination}
\begin{table}
 \centering
 \caption{K-point grids for the final phases of high-pressure fluorine}
 \begin{tabular}{c c}
  Phase & k-point grid\\
  \hline
  (12) $C2/m-2$ & $16 \times 16 \times 12$ \\
  (71) $Immm-2$ & $16 \times 16 \times 12$ \\
  (139) $I_{4}/mmm-2$ & $12 \times 16 \times 16$ \\
  (225) $Fm\bar{3}m-4$ & $12 \times 12 \times 12$ \\
  (12) $C2/m-6$ & $12 \times 16 \times 12$ \\
  (15) $C2/c-8$ & $4 \times 4 \times 4$ \\
  (64) $Cmca-8$ & $16 \times 16 \times 16$ \\
  (131) $P_{42}/mmc$ & $16 \times 16 \times 16$ \\
  (223) $Pm\bar{3}n$ & $16 \times 16 \times 16$ \\
 \end{tabular}\label{table2}
\end{table}

From USPEX predictions, 14 different crystal structures emerged as potential candidates for fluorine in the terapascal regime. These structures included all those previously reported that were compatible with our cell sizes (\textit{i.e.}, except for $P6/mcc$ which contains 14 atoms per cell, since this size was not included in our study) and also a triclinic form. An additional closer inspection demonstrated that some phases were very similar to each other: in those cases the ones with lower symmetry were, in fact, slightly distorted cells of those with higher symmetry, with differences of around 1\% in their cell parameters and atomic positions. This realization allowed us to narrow the list down to 9 structures: $C2/m-2$, $Immm-2$, $I_{4}/mmm-2$, $Fm\bar{3}m-4$,  $C2/m-6$,  $C2/c-8$, $Cmca-8$, $P_4{2}/mmc$, and $Pm\bar{3}n$. More detailed calculations were carried out for the final set of structures. For instance, additional physical properties were calculated using Quantum ESPRESSO with a projector-augmented wave pseudopotential including seven valence electrons. For all final calculations, the kinetic energy cutoff for the plane-wave basis set was 200 Ry and their correspondent k-point grid for the Brillouin zone integration provided a total energy difference convergence of 2 meV per atom or better. The details are displayed in table \ref{table2}.

\begin{figure}[ht]
\centering
\includegraphics[width=\linewidth]{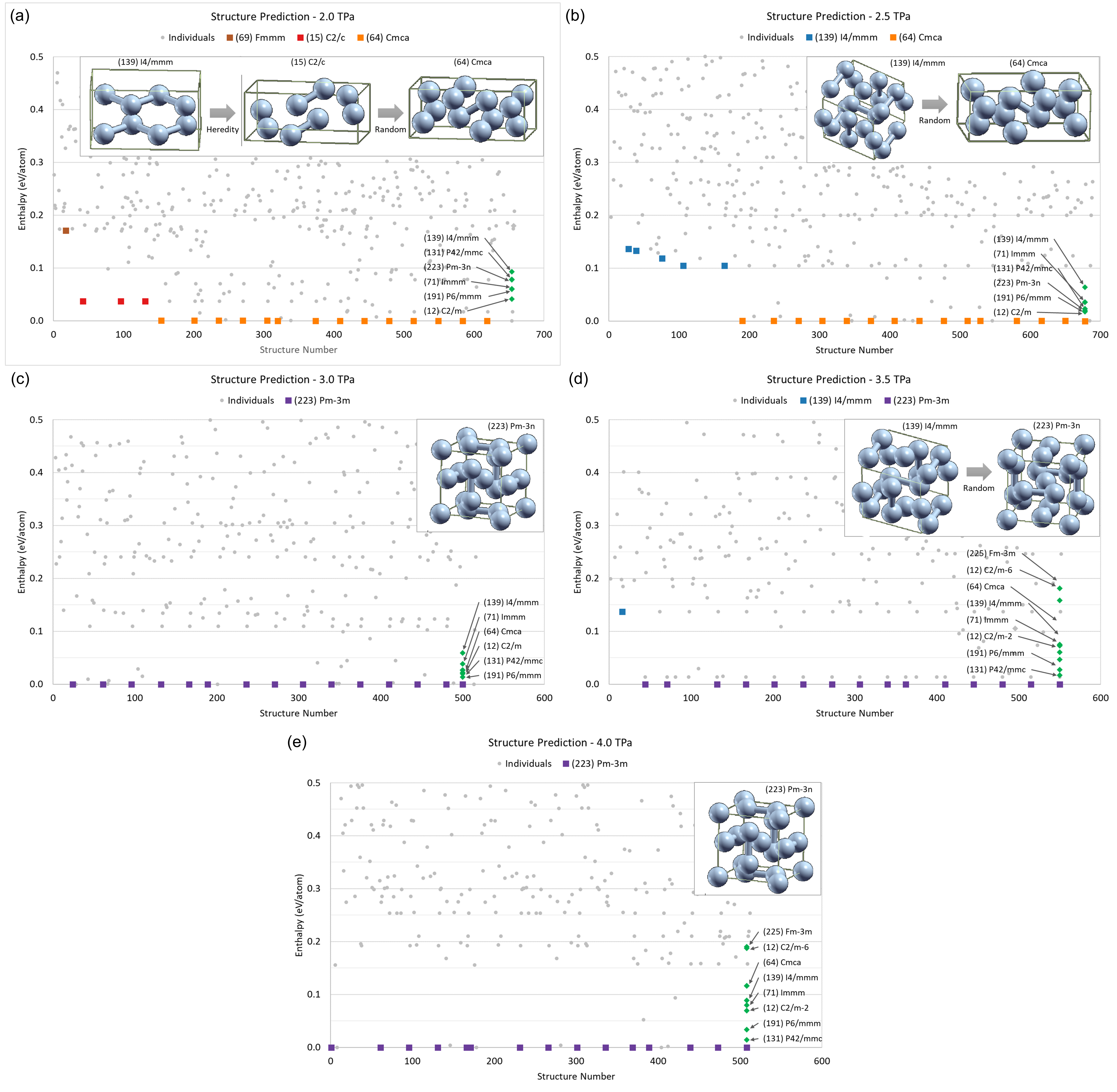}
\caption{Structure prediction process for fluorine with 8 atoms per cell within the pressure range from 2.0 to 4.0 TPa. Light gray circles exhibit the progress of each evolutionary simulation. The squares show the lowest enthalpy obtained at a given generation, with the colors indicating the space group of the best individuals: $C2/c$ (red), $Fmmm$ (brown), $Cmca$ (orange), $I4/mmm$ (blue), and $Pm\bar{3}n$ (purple). The insets show as 3D-images the sequences in the process of predicting the best structures by using the genetic algorithm implemented in USPEX.}
\label{figure2}
\end{figure}

Figure \ref{figure2} shows the case of 8 atoms per cell at selected pressures after variable-cell optimization from \textit{ab initio} results for the forces and cell stresses. USPEX was able to predict those phases proposed in previous works \cite{Olson-2020, Duan-2021} after just a few iterations. The insets show structures with the best individual enthalpy per atom at each generation; the squares mark the minimum enthalpy corresponding to different space groups:$C2/c$ (red), $Fmmm$ (brown), $Cmca$ (orange), $I4/mmm$ (blue), and $Pm\bar{3}n$ (purple). Each green rhombus indicates the relative enthalpy of other structures obtained with USPEX and reported in Refs. \cite{Olson-2020} and \cite{Duan-2021}.

\begin{figure}[ht]
\centering
\includegraphics[width=\linewidth]{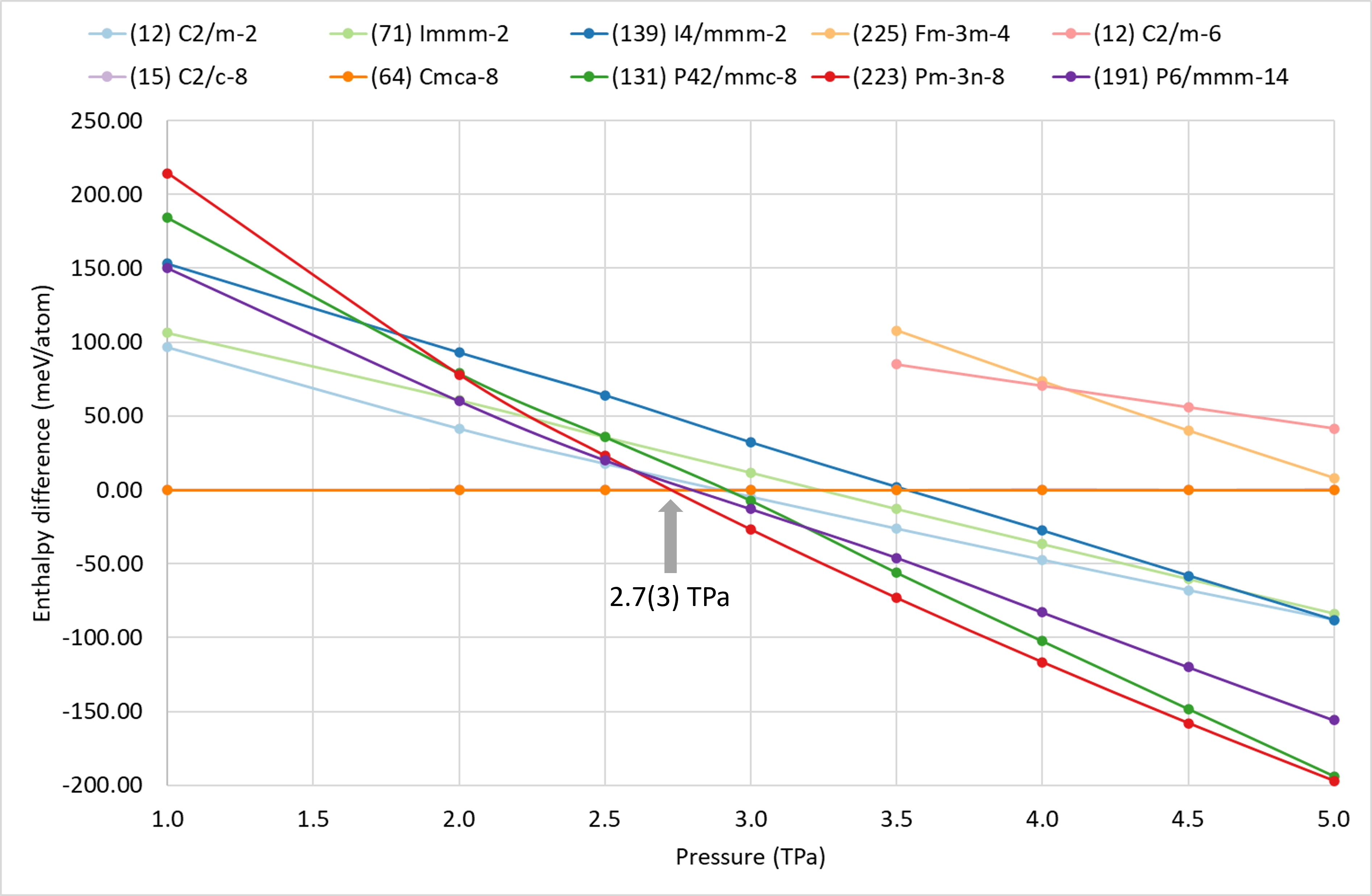}
\caption{Enthalpy relations of the structures found with the USPEX code and those reported in Refs. \cite{Olson-2020} and \cite{Duan-2021}, relative to the $Cmca$ structure as a function of pressure. The final number after the dash, in the name of each structure, represents the number of atoms per conventional unit cell.}
\label{figure3}
\end{figure}

In figure \ref{figure3}, the enthalpy relations for the structures found in this work, along with those previously proposed, are represented. The molecular phase $Cmca$ remains stable up to 2.7(3) TPa. This persistence of the phase $Cmca$ up to the terapascal regime was also reported in Ref. \cite{Duan-2021}, and our USPEX calculations confirm that statement. $Cmca$ transforms directly into $Pm\bar{3}n$, which was also reported as a stable form for solid fluorine by Olson \textit{et al.} \cite{Olson-2020} and Duan \textit{et al.} \cite{Duan-2021}. Nevertheless, in those two previous references the $Pm\bar{3}m$ phase emerges at a higher pressure and after a reported intermediate phase: $P4_{2}/mmc$ for Ref. \cite{Olson-2020} and $P6/mmc$ for Ref. \cite{Duan-2021}, which represent contradictory findings. In fact, our USPEX results found the phase $P4_{2}/mmc$ (green line in Fig. \ref{figure3}) which shows an interesting tendency, but its enthalpy is not the lowest at any pressure within the range of our current study, although the situation could change at higher pressures. On the other hand, since the $P6/mmc$ structure reported in Ref. \cite{Duan-2021} did not appear in the USPEX search because it has 14 atoms per cell, we included it manually in our calculations and its enthalpy is also plotted (violet line in Fig. \ref{figure3}). However, same as for $P4_{2}/mmc$, $P6/mmc$ also does not become the most stable within our pressure range, therefore, our results do not support the findings from previous works regarding the existence of an intermediate phase between $Cmca$ and $Pm\bar{3}n$ at zero temperature.

\begin{figure}[ht]
\centering
\includegraphics[width=\linewidth]{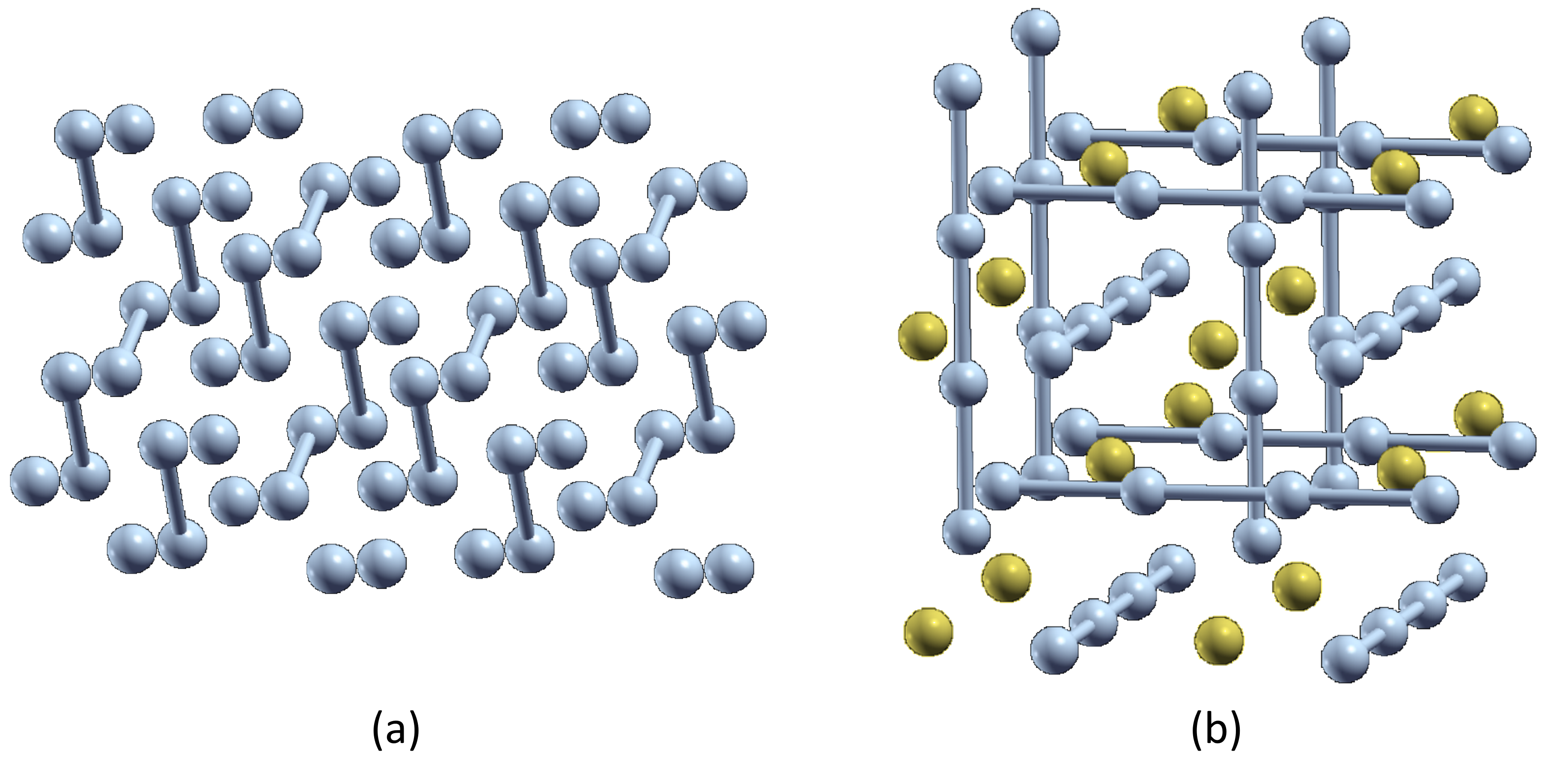}
\caption{Fluorine primitive cell of (a) molecular $Cmca$ at 1.0 TPa, and (b) polymeric-atomic $Pm\bar{3}n$ at 4.0 TPa consisting of linear chains (gray spheres) and atoms residing in the voids between chains (yellow).}
\label{figure4}
\end{figure}

Finally, regarding the two most stable structures found in this work, we can conclude tha $Cmca$ is a molecular phase, since, for example at 1.0 TPa, the nearest F--F distance is calculated as 1.28 \AA, and the second nearest F--F distance is 1.55 \AA, which clearly establishes the molecular character of this crystalline structure which is shown in Fig. \ref{figure4}(a). On the other hand, the phase $Pm\bar{3}n$ is composed of two types of F atoms as shown in Fig. \ref{figure4})(b): 75\% of them (in gray) are polymer linear (1D) chains that go along the lattice vectors, and the remaining 25\% (in yellow) are atoms located in the otherwise empty space between the linear chains, as shown in Fig. \ref{figure4})(b). The mixed polymeric-atomic F character in phase $Pm\bar{3}n$ is consistent with the reports from Refs. \cite{Olson-2020} and \cite{Duan-2021}. At 4.0 TPa, the distance between atoms within the polymeric chain is 1.26\AA, and the closest distance between an atom located in an interstice and an atom in a linear chain is 1.40\AA.

\section{Conclusions}

In summary, using the evolutionary approach implemented in the USPEX code and supported by DFT for calculating forces and stresses, we corroborated the existence of a transition from the widely stable $Cmca$ molecular form of fluorine towards a non-molecular phase that is composed of a mixture of linear chains and isolated atoms. So far, the stability of a pure element's molecular solid is one of the highest in fluorine, rivaled only by oxygen that also shares with fluorine its high electronegativity, confirming the expectation that these two elements should behave similarly in many ways among which also by being reluctant to become fully polymeric. Previous contradictory reports of intermediate stable phases existing in the transition between $Cmca$ and $Pm\bar{3}n$ were not supported here at zero temperature. Additionally, the phase changes reported here at high pressure may be the initial step towards further work on this element, \textit{i. e.}, calculating the finite temperature contributions to the free energy for determining how the relative stability of the different structures may change. Finally, this work may be the inspiration for exploring the behavior of other halogen elements at high pressure.

\section{Acknowledgements}

B. H. C.-O. thanks the financial support from MinCiencias under grant No. 80740-585-2021. J. A. M. thanks the Vicerrectoría de Investigaciones of the Universidad de Cartagena, for the support of the Grupo de Modelado Computacional through internal grants. The numerical calculations in this article were performed on the ROSMME High-Performance Computing Center of the University of Cartagena.

\bibliography{main}
\bibliographystyle{vancouver}

\end{document}